\documentclass[reprint,amsmath,amssymb,aps,pra]{revtex4-1}

\usepackage{graphicx}
\usepackage{epsfig}
\usepackage{epstopdf}
\usepackage{subfigure,color}
\usepackage{dcolumn}
\usepackage{bm}
\usepackage{lineno}

\def\e{\begin{equation}}
\def\f{\end{equation}}
\def\_#1{{\bf #1}}
\def\/#1{_{\rm #1}}
\def\.{\cdot}

\begin{document}

\preprint{APS/123-QED}
		
\title{Unlimited Accumulation of Electromagnetic Energy Using Time-Varying Reactive Elements}
	
\author{M.~S.~Mirmoosa, G.~A.~Ptitcyn, V.~S.~Asadchy and S.~A.~Tretyakov}
	
\affiliation{Department of Electronics and Nanoengineering, Aalto University, P.O.~Box 15500, FI-00076 Aalto, Finland}
	
\date{\today} 

\begin{abstract}
Accumulation of energy by reactive elements is limited by the amplitude of time-harmonic external sources. In the steady-state regime, all incident power is fully reflected back to the source, and the stored energy does not increase in time, although the external source continuously supplies energy. Here, we show that this claim is not true if the reactive element is time-varying, and time-varying lossless loads of a transmission line or lossless metasurfaces can accumulate  electromagnetic energy supplied by a time-harmonic source continuously in time without any theoretical limit. We analytically derive the required time dependence of  the load reactance and show that it can be in principle realized as a series connection of mixers and filters.  Furthermore, we prove that properly designing time-varying $LC$ circuits one can arbitrarily engineer the time dependence of the current in the circuit fed by a given time-harmonic source. As an example, we theoretically demonstrate a circuit with a linearly increasing current through the inductor. 
Such $LC$ circuits can accumulate huge energy from both the time-harmonic external source and the pump which works on varying the circuit elements in time. 
Finally, we discuss how this stored energy can be released in form of a time-compressed pulse. 
\end{abstract}
\maketitle

\section{Introduction}
\label{sec:Introduction}  
Recently, time-space modulated structures attracted significant interest especially in realizing magnetless nonreciprocal devices. However, most of the studies have been limited to time-harmonic modulations, extending the classical results on mixers and parametric amplifiers. In this paper we look at new possibilities which can open up if we allow \emph{arbitrary} time modulations of structure parameters (circuit reactances, material permittivity, etc). We expect that this approach may allow overcoming a number of important limitations inherent to conventional, harmonically modulated elements. As an interesting for applications example we consider capturing, accumulating, and storing of electromagnetic energy.

The use of conventional {\it lossless reactive elements} for energy accumulation is not efficient. Based on the circuit theory, we know that two fundamental reactive elements, the inductance and capacitance, store electromagnetic energy. If a time-varying current source $i(t)$ is connected to an inductor $L$, the stored  magnetic energy equals $W_{\rm{m}}(t)=Li(t)^2/2$. Replacing the inductor by a capacitor $C$ and employing a time-varying voltage source $v(t)$, we can store electric-field energy $W_{\rm{e}}(t)=Cv(t)^2/2$~\cite{Jackson_book}. Obviously, the stored energy is indeed limited by the source. To accumulate larger energy we need to have sources with larger output current or voltage. Even more importantly, most  practically available sources whose energy we can harvest provide time-periodical (in particular, time-harmonic) output. In the case of time-harmonic sources ($i(t),\,v(t)=A\cos(\omega t)$, where $\omega$ and $A$ represent the angular frequency and the amplitude, respectively), the stored energy is fluctuating between zero and $A^2L/2$ (inductor fed by a current source) or $A^2C/2$ (voltage-source fed capacitor). Therefore, the maximum amount of energy we can exploit is at $t=nT$, where $n=1,2,3,\dots$ ($T$ denotes the period), and it is completely determined by the amplitude of the external source. 

Now, the intriguing question which we consider here is \emph{if we can exceed this limitation for time-harmonic sources and continuously accumulate the energy supplied by the source in some reactive elements}. The fundamental problem here is to eliminate reflections from a reactive load, so that all the incident power will be accumulated in the load and made available at some desired moment of time. If it is possible to control the time dependence of the external source, this problem can be solved by making the external voltage or current exponentially growing in time~\cite{Baranov}. Here, we are interested in more practical scenarios when the external source cannot be controlled (for instance,  energy harvesting). Thus, we assume that the external source provides a given time-harmonic output and introduce solutions employing time-dependent reactive elements $L(t)$ and/or $C(t)$. Note that the discussion here equally applies to circuits, waveguides, or waves incident on lossless boundaries, because in each of these cases the reflection and absorption phenomena can be modeled using an equivalent  reactive load impedance. 

Generally speaking, the use of time-varying elements in electronic circuits~\cite{Cullen, Fettweis, Kamal, Currie, Baldwin, Macdonald, Anderson, Liou, Strom} as well as time-varying material properties~\cite{Slater, Simon, Hessel, Holberg, Gonzalez, Biancalana, Fan, Sounas, Fleury1, Fleury2, Alu} (usually, time-varying permittivity) for engineering wave propagation have been attracting the researchers' attention since approximately 1960s till today. These works are mainly focused on achieving  nonreciprocity, amplification and frequency conversion. Here, we use time-varying reactive elements for energy accumulation. While in most of the earlier studies,  periodical time variations of circuit elements have been used, for our goals we will need to consider arbitrary time variations of parametric elements. 

The paper is organized as follows: In Section~\ref{sec:zreftmi}, we apply the transmission-line theory and find the required condition to have zero reflection (unlimited accumulation of energy) in a line that is terminated by a single time-modulated reactive load. Subsequently, we elucidate our  result by drawing an analogy between our time-modulated load and two different scenarios explained in the parts A and B of Section~\ref{sec:zreftmi}, respectively. In Section~\ref{sec:opaopaopapa}, we go one step further and consider a load which comprises two time-varying reactive elements that are in parallel with each other. In contrast to the case studied in Section~\ref{sec:zreftmi}, here we show that it is possible to engineer the electric currents flowing through the elements while we still obtain zero reflection. We explain how engineering the electric currents affects the amount of energy accumulated by the entire load. Finally, Section~\ref{sectionconclulast} concludes the paper. This paper is our first step that hopefully opens a way for further work on systems with arbitrary time modulations of parameters and, in particular, for practical investigations of efficient devices which accumulate energy from time-periodical, low-amplitude sources. We expect that the use of non-harmonic time modulations of system parameters will offer other, more general means to shape system response at will.     

\section{Zero-reflection from time-modulated reactive elements}
\label{sec:zreftmi}
\begin{figure*}[t!]\centering
	\includegraphics[width=15cm]{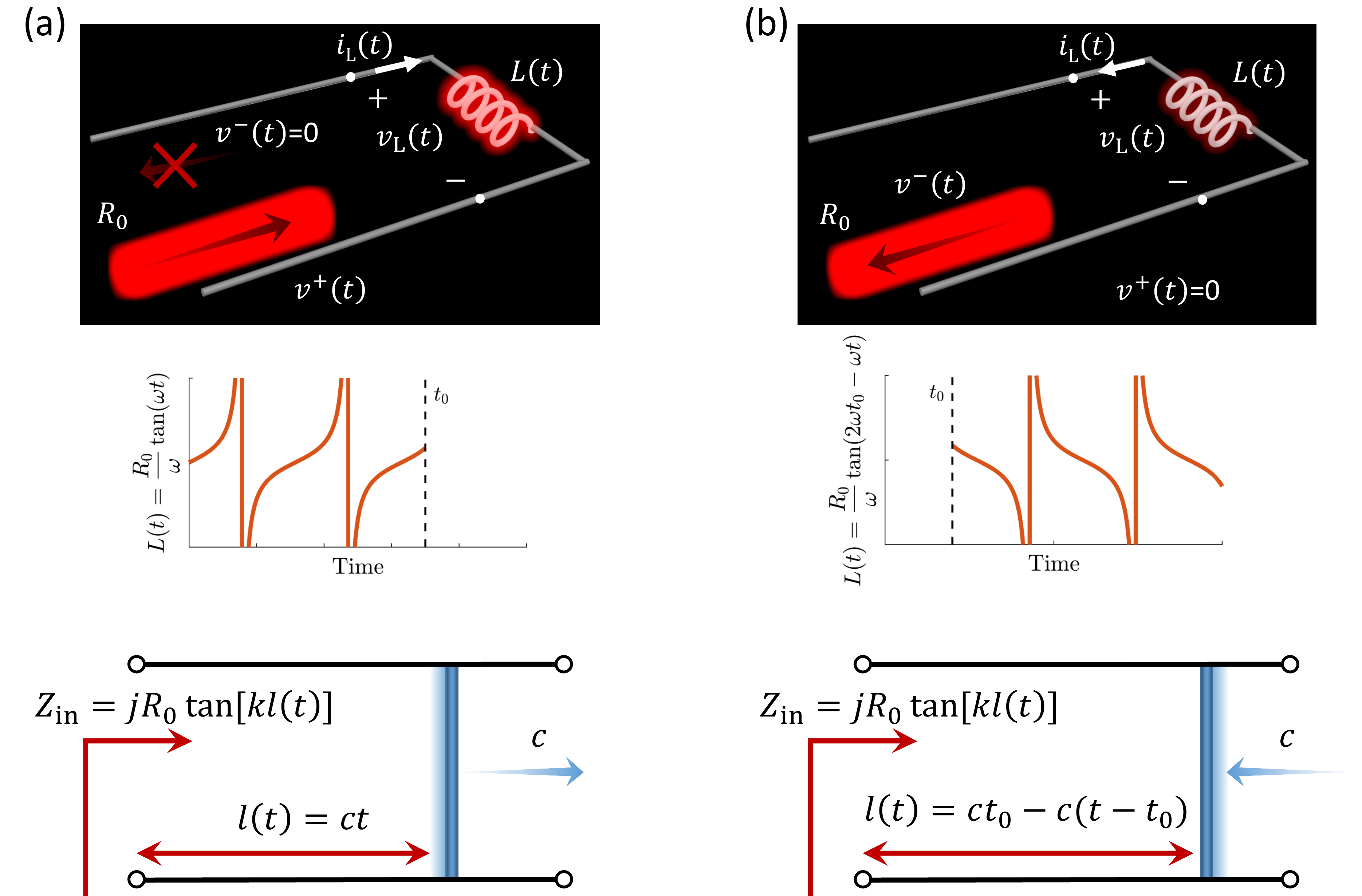}
	\caption{(a)--Transmission line terminated by a time-dependent inductance which absorbs the incoming energy. The bottom illustration shows its conceptual realization by a short-circuited line whose length extends with a constant velocity. (b)--Energy releasing by the same line but with different inductance modulation (inverse with respect to moment $t_0$). This means that in its conceptual realization, the load is moving in the opposite direction.} 
	\label{fig:parallel}
\end{figure*}
For a linear and time-invariant transmission line fed by a time-harmonic voltage/current source, the supplied energy is fully transferred to the load if the characteristic impedance of the line is equal to the load impedance of that line~\cite{Pozar}. Apparently, for the case of a lossless line terminated by a conventional inductance or capacitance, the energy is entirely reflected back as the incident wave arrives at the load, and therefore, there is no absorption nor accumulation of energy in the load. Mathematically, this property follows from the fact that the characteristic impedance of the line is real while the load impedance is purely imaginary (reactive). 

It is easy to see that making the load reactance vary in time, it is possible to emulate resistance, although there is no actual power dissipation or generation. Indeed, the voltage over a capacitor/inductor and the current flowing through the element are related to each other as
\begin{equation}
\begin{split}
&v_{\rm{L}}(t)=L(t){di_{\rm{L}}(t)\over dt}+{dL(t)\over dt}i_{\rm{L}}(t),\cr
&i_{\rm{C}}(t)=C(t){dv_{\rm{C}}(t)\over dt}+{dC(t)\over dt}v_{\rm{C}}(t),
\end{split}
\label{eq:LC}
\end{equation}    
where  $v_{\rm{L,C}}(t)$ and $i_{\rm{L,C}}(t)$ denote the instantaneous voltage and current in a time-varying inductor or capacitor, respectively. Conventionally, when the element is time-independent, the second term in the above equations (${dL(t)\over dt}$ or ${dC(t)\over dt}$) vanishes and therefore, the voltage and current are proportional by a factor which is purely imaginary in the frequency domain. However, the scenario is completely different as the inductance/capacitance element varies with respect to time. The second term is not zero any more and it has the form of the usual Ohm law, where the role of the resistance or conductance is played by the time derivatives of the circuit reactances. Clearly, this virtual resistance or conductance describes virtual absorption of energy, which can be actually accumulated in the reactive element. Next, we will study how this possibility can be exploited for accepting  and accumulating incident energy in reactive elements. 

\begin{figure}[t!]\centering
\subfigure[]{\includegraphics[width=0.4\textwidth]{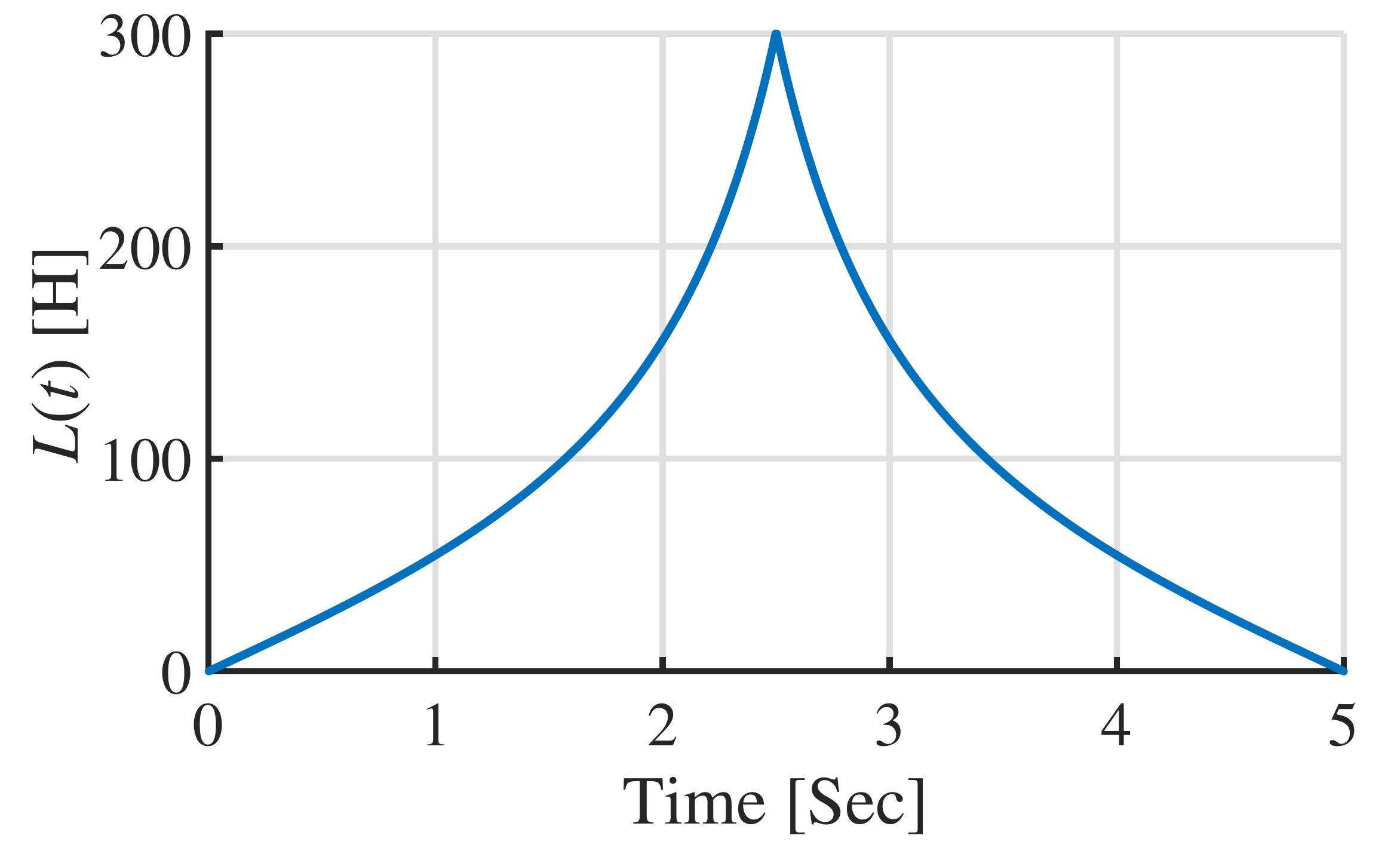}}
\subfigure[]{\includegraphics[width=0.4\textwidth]{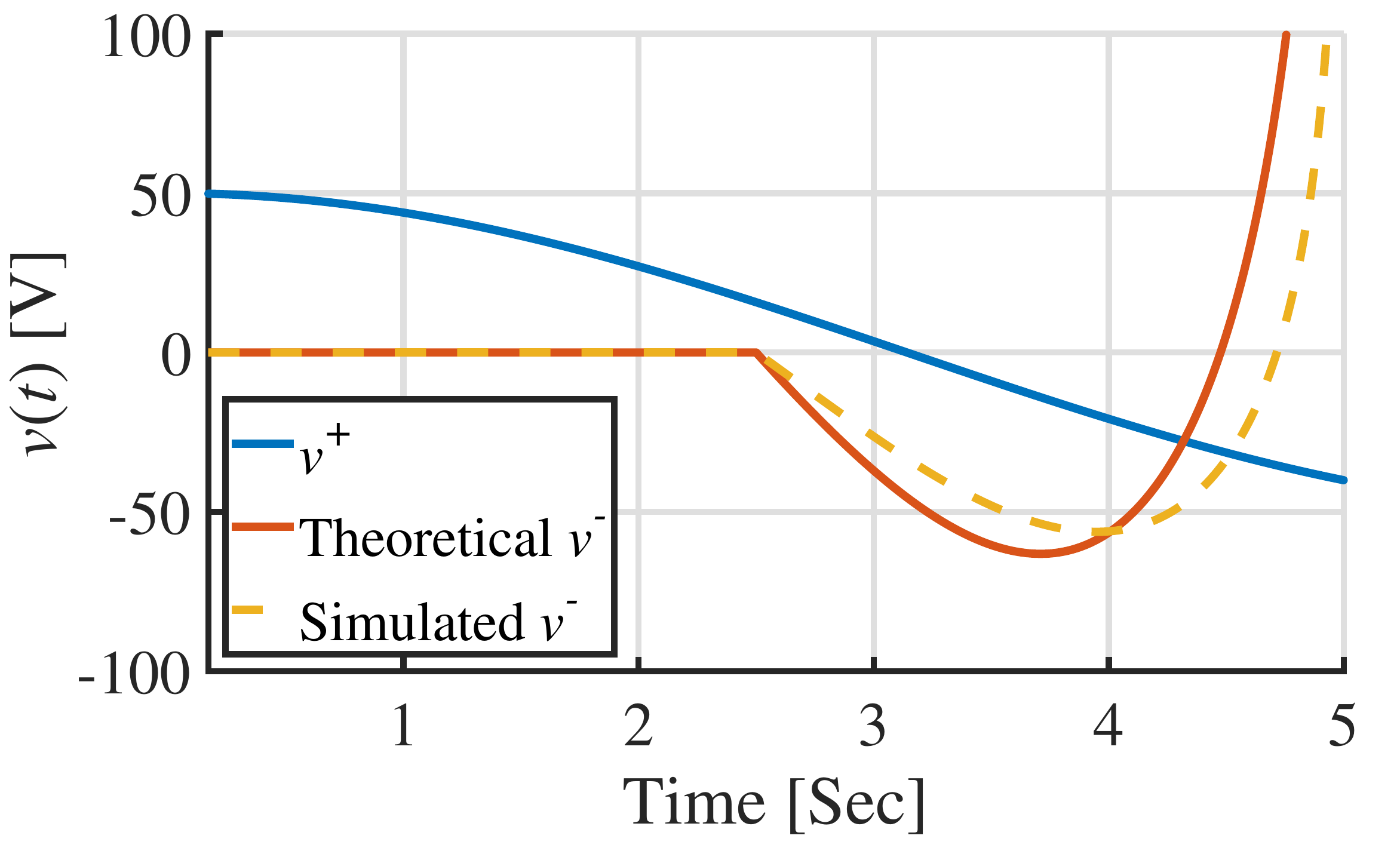}}
\caption{(a)--Time-modulated inductance, as the load of the transmission line, in the accumulation and releasing regimes. (b)--The incident and reflected waves (simulated and theoretical) at the load position. Here, notice that $R_0=50\,\Omega$ and $v^+(t)=50\cos(\omega t)$ V where $\omega=0.5$ rad/sec.}
\label{fig:sarefwav}
\end{figure}

Let us consider a lossless transmission line loaded by a time-varying inductance $L(t)$ and denote the voltage of the wave propagating towards the load as $v^+(t)$. The reflected voltage wave is denoted by $v^-(t)$. The instantaneous voltage over the load and the current flowing through the load are written as
\begin{equation}
\begin{split}
&v_{\rm{L}}(t)=v^+(t)+v^-(t),\cr
&i_{\rm{L}}(t)={v^+(t)-v^-(t)\over R_0},
\end{split}
\label{eq:fbwardwaves}
\end{equation}
where $R_0$ represents the characteristic impedance of the line. On the other hand, the voltage and the current are related to each other by Eq.~(\ref{eq:LC}). Substituting Eq.~(\ref{eq:LC}) into Eq.~(\ref{eq:fbwardwaves}), we derive a general formula for the incoming and reflected waves at the load position as
\begin{equation}
\begin{split}
&L(t){dv^-(t)\over dt}+\bigg[{dL(t)\over dt}+R_0\bigg]v^-(t)=\cr
&L(t){dv^+(t)\over dt}+\bigg[{dL(t)\over dt}-R_0\bigg]v^+(t).
\end{split}
\label{eq:reflwave}
\end{equation}  
In this equation, the left-hand side contains the terms measuring the reflected wave, while the right-hand side depends on the incident voltage $v^+(t)$ only. For a given incident voltage  $v^+(t)$ and any time-dependent inductance $L(t)$  we can find the reflected wave by solving the above first-order differential equation. Using Eq.~(\ref{eq:reflwave})  we can find ways to manipulate the reflected  wave by choosing the proper function for $L(t)$. Here, we are interested in accumulating all the incident energy, that is, we are interested in reactive loads which do not reflect. 
   
Now, assume that the incoming signal is time-harmonic, written as  $v^+(t)=A\cos(\omega t)$. In the following, the amplitude and the initial phase of the incident wave are supposed to be unity and zero, respectively, just for simplicity of formulas. Requiring absence of reflection ($v^-(t)=0$), we find the corresponding time dependence of  $L(t)$ from Eq.~(\ref{eq:reflwave}). Here, it is worth noting that the time-harmonic current in the load is in phase with the incident  voltage in case of zero reflection, as is obvious from Eq.~(\ref{eq:fbwardwaves}). The result reads 
\begin{equation}
L(t)={R_0\over\omega}\tan(\omega t).
\label{eq:TDinductance}
\end{equation}
Such inductance as a load virtually absorbs all the input energy. Similar considerations can be made for loads in form of time-varying capacitance or other reactive loads. In the following, we explain the reason why the time-varying inductance described by Eq.~(\ref{eq:TDinductance}) gives rise to the virtual absorption. To do this, we make an analogy between the inductance and a short-circuited line whose length is linearly increasing versus time. In addition, we also draw another analogy between the inductance and a load consisting of an infinite number of sinusoidal inductances connected in series.

\subsection{Short circuited line}
To understand the physical meaning of the result obtained above, we can notice a  similarity of this formula (Eq.~(\ref{eq:TDinductance})) with the classical formula for the input reactance of a short-circuited transmission line:
\begin{equation}
Z_{\rm{in}}=jR_0\tan(k l), \label{eq:zinp}
\end{equation}
where  $k={\omega/c}$ ($c$ denotes the phase velocity) is the phase constant and $l$ represents the length of the line. 
Obviously, the time-varying inductance $L(t)$ given by (\ref{eq:TDinductance}) is the same as that seen at the input port of a lossless short-circuited transmission line if the length of the transmission line is \emph{linearly increasing} with the constant velocity equal to $c$, since in this case $Z_{\rm{in}}=jR_0\tan(k l)=j\omega L(t)$ (see Fig.~\ref{fig:parallel}(a)). We see that in this conceptual scenario the reason for having no reflection from a lossless load is that the incident wave never reaches the reflecting termination, since the short is moving away from the input port with the same velocity as the phase front of the incident wave. Thus, varying the load inductance as prescribed by (\ref{eq:TDinductance}) for enough long time one can accumulate theoretically unlimited field energy in the reactive load. 

Let us now assume that we would like to release the stored energy. To do that, we would reverse the direction of the velocity of the short, i.e.~moving towards the input port (see Fig.~\ref{fig:parallel}(b)). All the energy stored in the line volume will be compressed in time and released into the feeding line at the moment when the length of the short circuited line becomes zero. The reactance of the line would correspond to a time-varying inductance given by 
\begin{equation}
L(t)={R_0\over\omega}\tan(2\omega t_0-\omega t),\,\,\,\,\,\,\,\,t_0<t<2t_0,
\label{eq:TDinductance_mirrored}
\end{equation}	   
in which $t_0$ is the moment when the velocity of the short changes the direction and consequently, the short moves backward. Based on our analogy between the time-varying inductance and the short-circuited line whose length changes versus time, we conclude that the time-modulated inductance is expressed as
\begin{equation}
\begin{split}
&L_{\rm{A}}(t)={R_0\over\omega}\tan(\omega t),\,\,\,\,\,\,\,\,0<t<t_0,\cr
&L_{\rm{R}}(t)={R_0\over\omega}\tan(2\omega t_0-\omega t),\,\,\,\,\,\,\,\,t_0<t<2t_0.
\end{split}
\label{eq:lllltttt}
\end{equation}
Here, $L_{\rm{A}}(t)$ and $L_{\rm{R}}(t)$ represent the time-modulated inductances in accumulation and releasing regimes, respectively. It is clear that the function describing $L_{\rm{R}}(t)$ is the mirror of the function describing $L_{\rm{A}}(t)$ with respect to the moment $t=t_0$.

We have simulated the system illustrated by Fig.~\ref{fig:parallel}(a) (top one)  utilizing  MathWorks Simulink software. We assume that our system accumulates the energy till the moment $t_0=2.5$~s and subsequently, it releases the energy till the moment $t=2t_0=5$~s. The modulation function for the reactive element expressed by Eq.~(\ref{eq:lllltttt}) is shown in Fig.~\ref{fig:sarefwav}(a), and the simulated and theoretical results for the reflected wave are shown in Fig.~\ref{fig:sarefwav}(b). Notice that the theoretical formula for the reflected wave can be obtained using Eq.~(\ref{eq:reflwave}). The blue color in Fig.~\ref{fig:sarefwav}(b) corresponds to the incident wave [$v^+(t)=50\cos(\omega t)$ V] while the red/yellow color represents the reflected wave. As it is seen, the reflection is zero till $t_0=2.5$~s meaning that the reactive element stores the electromagnetic energy (virtual absorption). After $t=t_0$, the reflection appears and we are in the releasing regime. The theoretical and simulated results for the reflected wave are in good agreement.

Let us next consider the effects of inevitable dissipation losses in the system.  Concentrating  on the accumulating regime, we add a resistance ($R_{\rm{L}}$) to the load in order to see the effect of ohmic losses. This load resistance is  connected in series with the time-varying inductance. We have simulated again the structure in Fig.~\ref{fig:parallel}(a) and observed that if the load resistance is smaller than about $2$\% of the characteristic impedance of the transmission line (1~Ohm for our example of a 50-Ohm line), the reflected wave is still near zero and the system works quite well. For resistances larger than 1~Ohm ($R_{\rm{L}}>1\Omega$), some  reflection appears and the efficiency decreases.

Interestingly, it is in fact possible to eliminate any reflections also for lossy terminations (with arbitrary values of $R_{\rm{L}}$) by a simple modification of the modulation function of the time-varying inductance $L_{\rm{A}}(t)$. This way we can compensate the resistive losses completely. Indeed, if we rewrite Eq.~(\ref{eq:reflwave}) by assuming that there is also the resistance $R_{\rm{L}}$ at the termination, we see that the required time-varying inductance to obtain zero reflection becomes 
\begin{equation}
L_{\rm{A}}(t)={R_0\over\omega}\left(1-{R_{\rm{L}}\over R_0}\right)\tan(\omega t),\,\,\,\,\,\,\,\,0<t<t_0.
\label{eq:modmodLATRL}
\end{equation}
According to the above expression, if $R_{\rm{L}}=0$, we will achieve the same result as given by Eqs.~(\ref{eq:TDinductance}) or (\ref{eq:lllltttt}). Note that if $R_{\rm{L}}=R_0$, then the modulated  inductance needed to eliminate reflections is zero because in this limiting case  the load is already  perfectly matched to the transmission line (perfect absorption condition). Our simulations have confirmed that the modification of the modulation function given by Eq.~(\ref{eq:modmodLATRL}) indeed results in zero reflection in the accumulating regime for different values of loss resistance $R_{\rm{L}}$.

\begin{figure}[t!]\centering
	\includegraphics[width=0.5\textwidth]{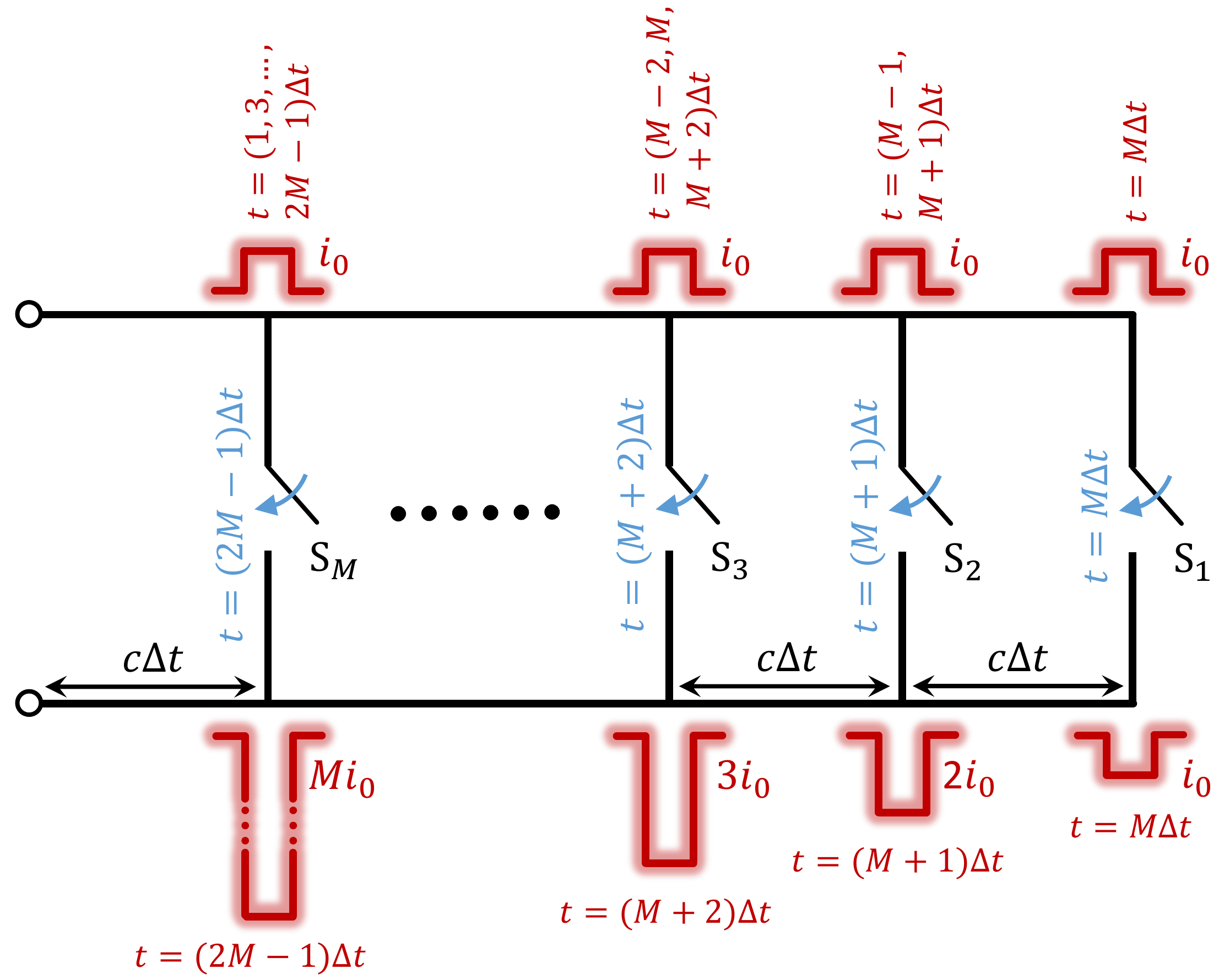}
	\caption{Space-time modulated  transmission line with $M$ switches. Each switch is closed at specific time moments shown in blue color. Amplitudes of the signal at different locations of the transmission line at different time moments are depicted in red color.}
	\label{fig:Switches}
\end{figure}

Similar effects of energy accumulation and release can be achieved using a transmission line periodically loaded with switches which can be switched at the appropriate time moments, as required by Eq.~(\ref{eq:zinp}). Figure~\ref{fig:Switches} schematically shows realization of this concept. Let us consider an incident signal in form of periodic pulses of amplitudes $i_0$ and durations $\Delta t$.
Prior to time  moment $t=M \Delta t$, the pulses enter the transmission line and propagate 
along the line  without any reflections because all the switches are open. 
 When the first (leading) pulse approaches switch~$S_1$ at time $t=M \Delta t$, it closes and the pulse reflects and starts propagating  in the opposite direction. Due to the properly adjusted   distance between the switches, at moment  $t=(M+1) \Delta t$, the first pulse is summed up (constructive interference) with the second pulse which has been reflected by switch~$S_2$. The amplitude of the resulting pulse is doubled: $2 i_0$. 
Likewise, at the position of each next switch the leading pulse amplitude is increased by $i_0$, resulting in the output pulse with amplitude $M i_0$.  In this scenario, the total energy entered the transmission line, proportional to $M i_0^2$, is compressed into a single pulse  with energy $(M i_0)^2$. This is not a passive design since, due to the energy conservation, extra work  proportional to $(M^2-M) i_0^2$ is required for  closing the switches.

\subsection{Load consisting of sinusoidal inductances}

\begin{figure*}[t!]\centering
	\includegraphics[width=12cm]{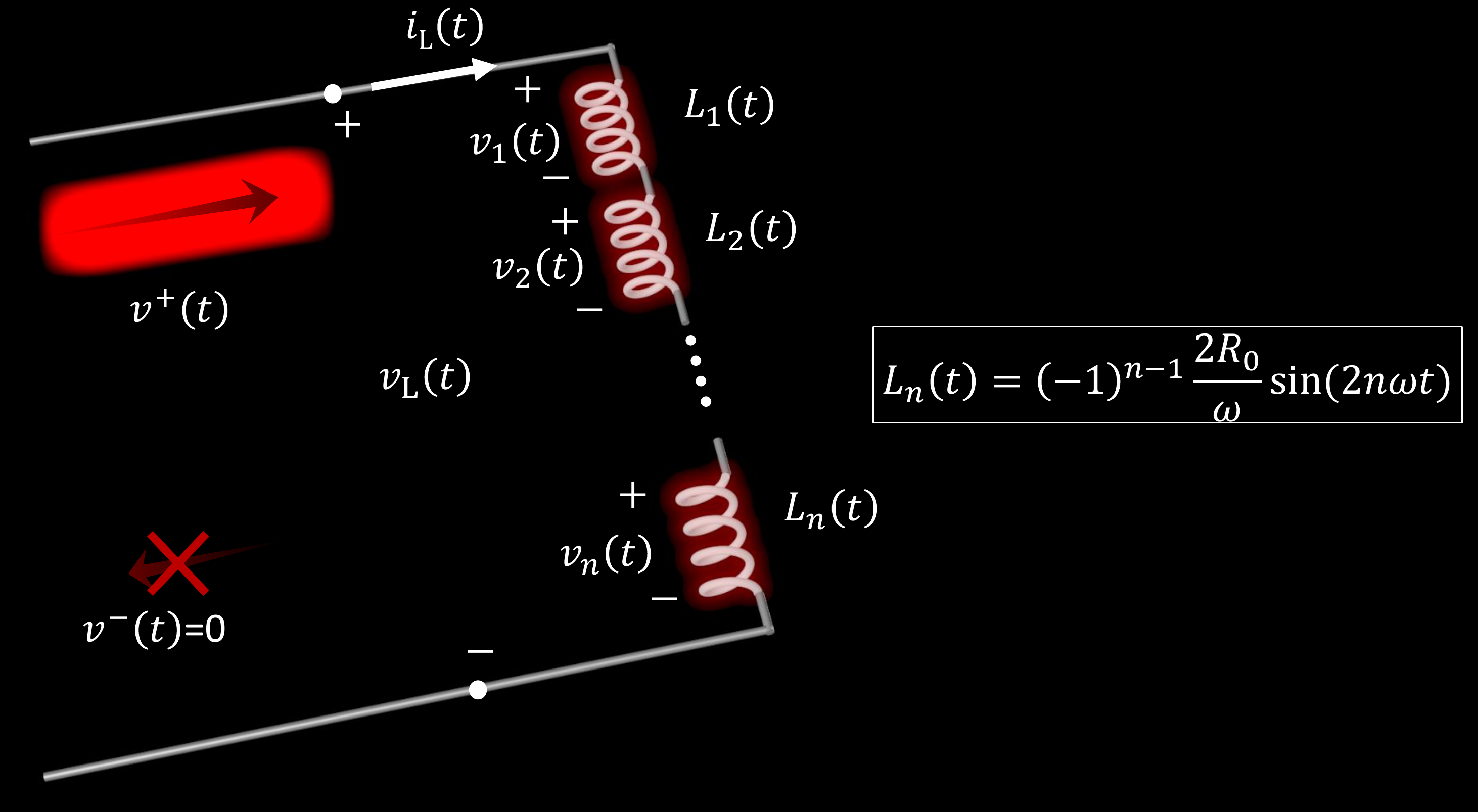}
	\caption{Transmission line terminated by an infinite number of time-dependent inductances connected in series.} 
	\label{fig:inducseries}
\end{figure*}

Another way to understand why the time-modulated inductance given by Eq.~(\ref{eq:TDinductance}) ensures full  virtual absorption is applying the Fourier series analysis. Since the required time-dependent load inductance given by Eq.~(\ref{eq:TDinductance}) is a periodical function, we can expand it in the Fourier series. We know that 
\begin{equation}
\tan(x)=2\sum_{n=1}(-1)^{n-1}\sin(2nx).
\end{equation}
Using the above equation, we can consider the time-dependent inductance as an infinite collection of harmonically modulated inductances which are connected in series, as is illustrated in Fig.~\ref{fig:inducseries}. Let us assume that the voltage over the whole load in the figure is $v_{\rm{L}}(t)=A\cos(\omega t)$ and the current flowing through the load is $i_{\rm{L}}(t)=A\cos(\omega t)/R_0$. Based on the Kirchhoff voltage law, the voltage $v_{\rm{L}}(t)=\sum_{n=1}v_n(t)$, where $v_n(t)$ is the voltage over each time-dependent inductance. Therefore,
\begin{equation}
v_n(t)=L_n(t){di_{\rm{L}}(t)\over dt}+{dL_n(t)\over dt}i_{\rm{L}}(t),
\label{eq:vnt}
\end{equation}
in which $L_n(t)=2(-1)^{n-1}{R_0\over\omega}\sin(2n\omega t)$. Simplifying Eq.~(\ref{eq:vnt}), we find that
\begin{equation}
\begin{split}
v_n(t)=A(-1)^{n-1}&\bigg[(2n-1)\cos((2n-1)\omega t)+\cr
&(2n+1)\cos((2n+1)\omega t)\bigg].
\end{split}
\label{eq:vntharmonics}
\end{equation}
Equation~(\ref{eq:vntharmonics}) shows that the $n$th time-dependent inductance operates as a mixer in which the input of this mixer is a time-harmonic current signal of the frequency $\omega$ having the amplitude equal to $A/R_0$ producing as the output two time-harmonic voltage signals of frequencies $(2n\pm 1)\omega$ and the amplitude of $(2n\pm 1)A$. The output signal can be amplified or attenuated depending on the integer number of the inductance element (it is amplified if $2n\pm 1>{1\over R_0}$). 

By substituting $n=1,\,2,\,3,...$ in Eq.~(\ref{eq:vntharmonics}), we realize that only the first harmonic corresponding to $n=1$ is not canceled in the series $v_{\rm{L}}(t)=\sum_{n=1}v_n(t)$ (which in the usual sense does not  convergence). The second term of $v_1(t)=A[\cos(\omega t)+3\cos(3\omega t)]$ cancels out with the first term of $v_2(t)=A[-3\cos(3\omega t)-5\cos(5\omega t)]$, and the second term of $v_2(t)$ is canceled by the first term of $v_3(t)$, and so on. Hence, only the first term $A\cos(\omega t)$ of $v_1(t)$ survives. Since the amplitude of this term is equal to the amplitude of the total voltage $v_{\rm{L}}(t)$,  the reflection coefficient equals zero. Here, it is worth mentioning that if we have a finite number of the time-dependent inductances shown in the figure, we can still emulate full absorption. From the above considerations we see that only the first harmonic $\omega$ and the harmonic $(2n+1)\omega$ remain. The other harmonics automatically vanish. Thus, to emulate full absorption, we only need to remove the $(2n+1)\omega$ harmonic by using a low-pass filter. If we do not filter this harmonic, certainly, the reflection is not zero.


\section{Time-dependent parallel $LC$ circuit}
\label{sec:opaopaopapa}

In the previous scenario with one reactive element, the electric current was limited by the characteristic impedance of the line and the amplitude of the incoming wave ($i(t)=A\cos(\omega t)/R_0$). The intriguing question is if {\it it is possible to realize any (growing) function for the electric current} flowing through the time-varying reactive element. Next we will show that it is indeed possible if the time-varying reactive load contains at least two reactive elements,  one inductive and one capacitive.  Having two connected circuit elements we have an additional degree of freedom to shape the electric current flowing through these components since (assuming parallel connection) only the sum of the two  currents should be equal to $i(t)=A\cos(\omega t)/R_0$ in order to ensure zero reflection. In this section we discuss the  design of such circuits and investigate the stored energy in the system. 

\begin{figure}[t!]\centering
	\subfigure[]{\includegraphics[width=0.45\textwidth]{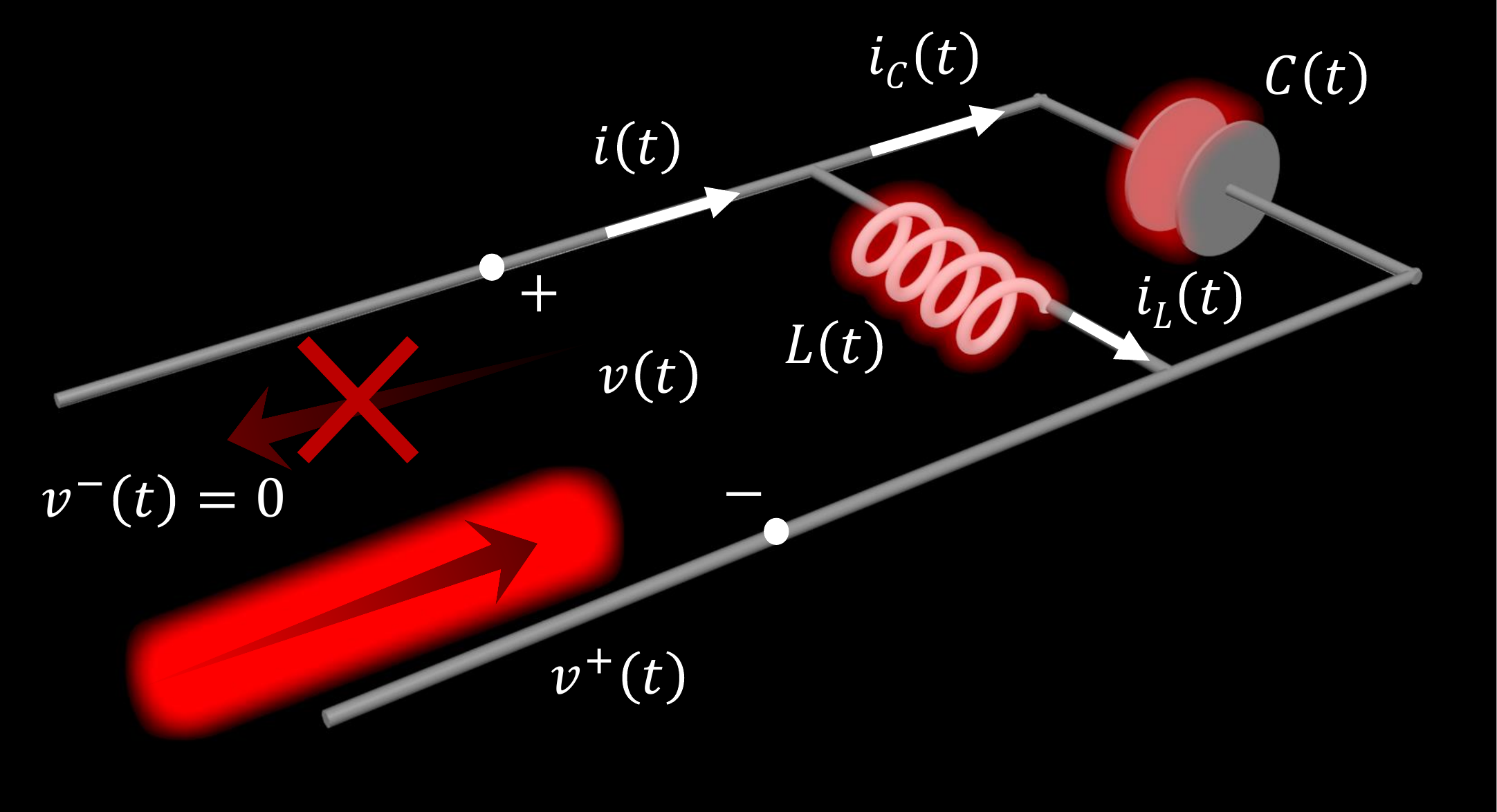}}
	\subfigure[]{\includegraphics[width=0.4\textwidth]{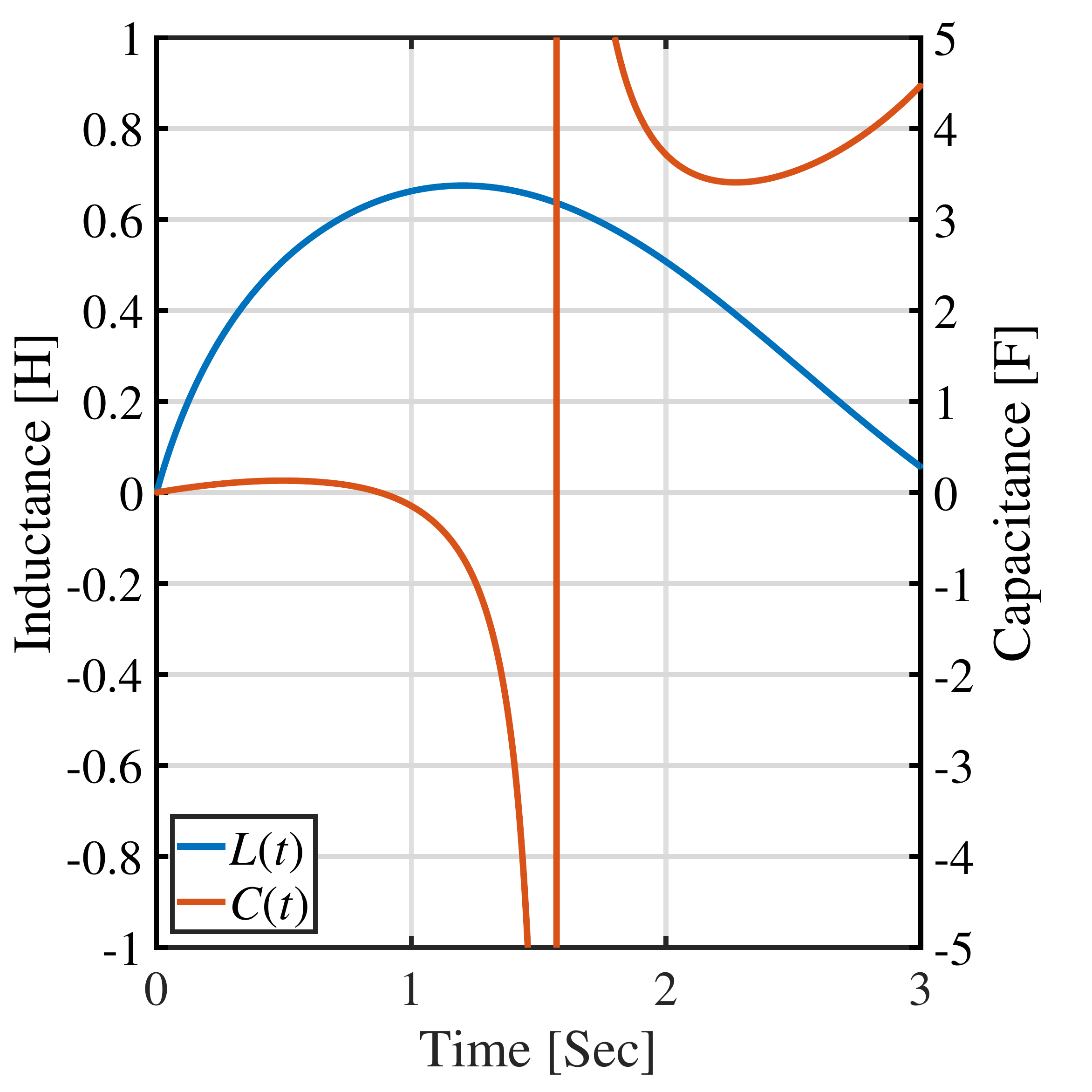}}
	\caption{(a) Transmission line terminated by a parallel $L(t)C(t)$-circuit. (b) Values of $L(t)$ and $C(t)$ required for realization of zero reflection regime.} 
	\label{fig:LC-circuit}
\end{figure}

Let us consider a transmission line terminated by a parallel $LC$-circuit which is formed by time-dependent components~$L(t)$ and~$C(t)$. Schematic of the circuit is illustrated by Fig.~\ref{fig:LC-circuit}(a). 
Suppose that the incident voltage wave is $v^+(t)=A\cos(\omega t)$ and the total electric current is $i(t)=A\cos(\omega t)/R_0$ (no reflection). Here, $i_{\rm L}(t)$ denotes  the current through the inductance and $i_{\rm C}(t)$ denotes the current through the capacitance. Based on Kirchhoff's current law, $i(t)=i_{\rm L}(t)+i_{\rm C}(t)$. This condition is fulfilled by setting the currents as
\begin{equation}
\begin{split}
&i_{\rm L}(t)={A\cos(\omega t)\over2R_0}+f(t),\cr
&i_{\rm C}(t)={A\cos(\omega t)\over2R_0}-f(t),
\end{split}
\label{eq:iLC_law}
\end{equation}
in which $f(t)$ can be an arbitrary function of time. As an example, we consider $f(t)$ as a linearly growing function  $f(t)=I_0t$ in which $I_0>0$ (this is only due to the simplicity of the function, here any differentiable function can be assumed). Applying Kirchhoff's laws and using Eq.~(\ref{eq:iLC_law}), we can find the required time dependences of the circuit elements. After some algebraic manipulations, we obtain the following expressions: 
\begin{equation}
\begin{split}
&L(t)=2R_0{A\sin(\omega t)\over\omega\bigg(A\cos(\omega t)+2R_0I_0t\bigg)},\cr
&C(t)=\frac{\tan(\omega t)}{2R_0\omega}-\frac{I_0t^2}{2A\cos(\omega t)}.
\end{split}
\label{eq:LtCt}
\end{equation}
As the above equation shows, the capacitance always possess asymptotes due to the tangent function. However, depending on the ratio between $R_0I_0$ and the angular frequency $\omega$ ($R_0I_0/\omega$), the inductance can be finite without having an asymptote. For example, Fig.~\ref{fig:LC-circuit}(b) shows the functions of $L(t)$ and~$C(t)$ for  $R_0=1$~$\Omega$, $\omega=1$~rad/sec, $A=1$~V and $I_0=1$~A/sec. At the initial moment, both elements are positive and growing. However, later the inductance decreases and goes to zero fluctuating around it due to the term  $\sin(\omega t)$ in the numerator.  

\begin{figure}[t!]
\centering
\subfigure[]{\includegraphics[width=0.42\textwidth]{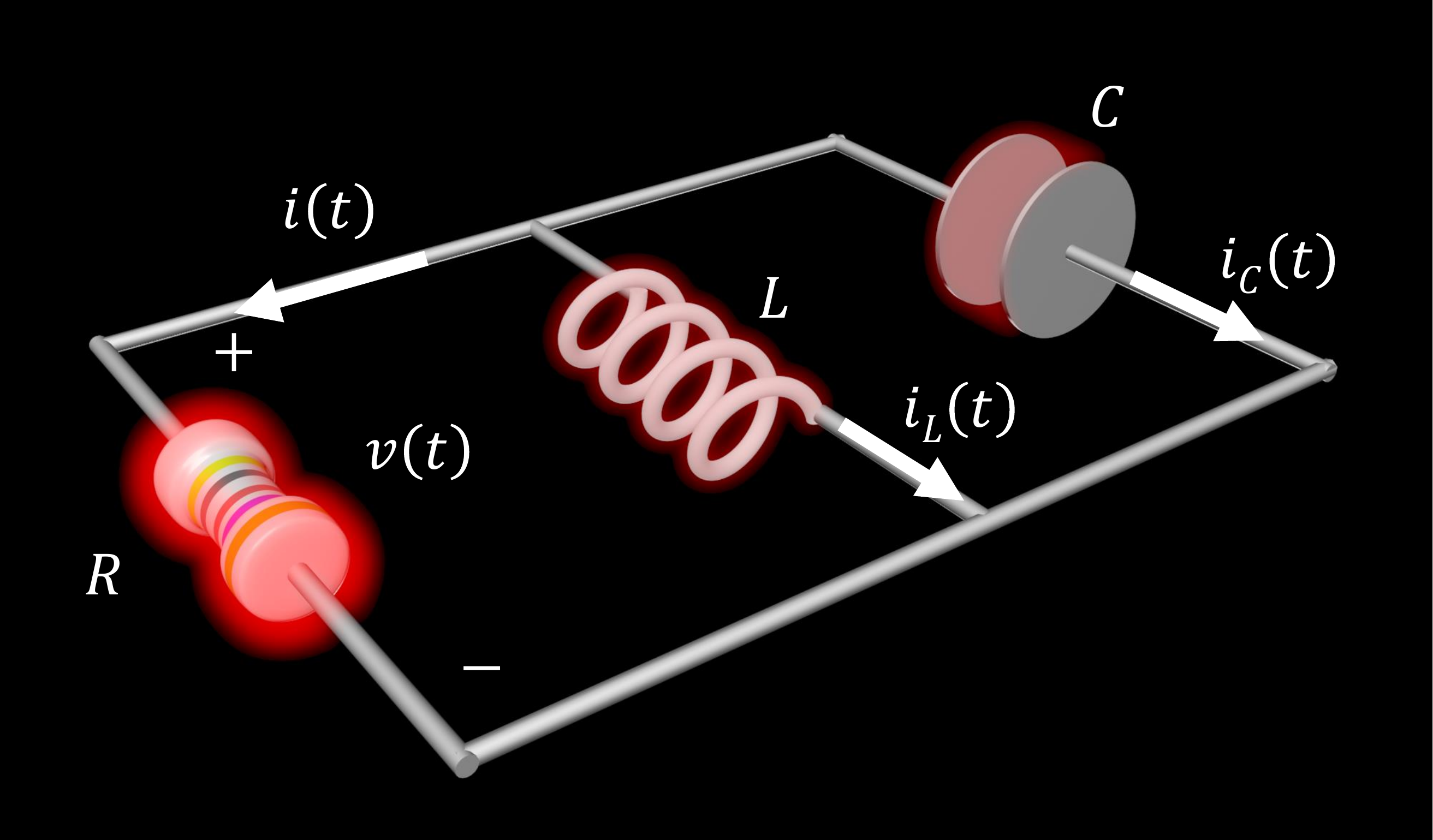}}
\subfigure[]{\includegraphics[width=0.4\textwidth]{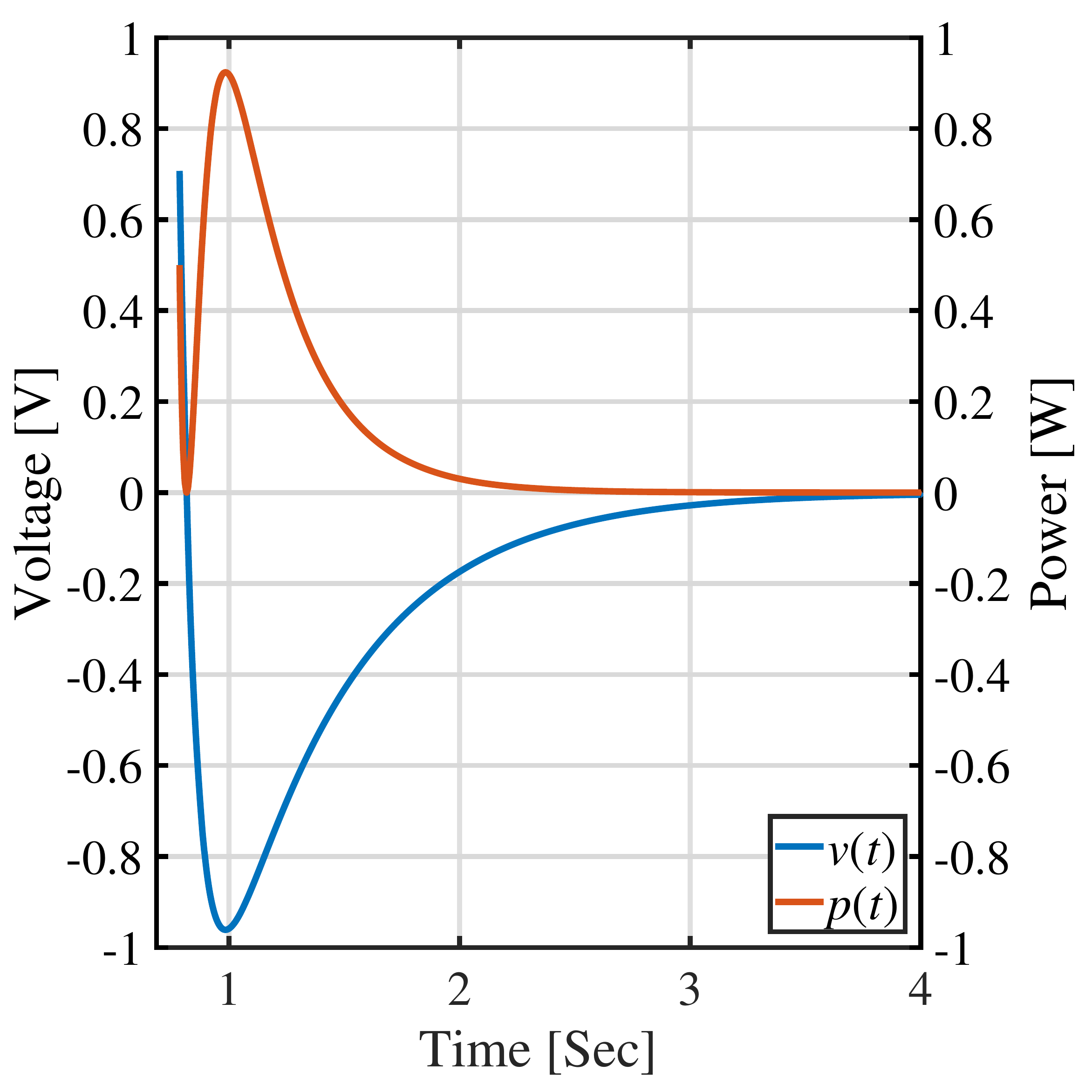}}
\caption{(a)--The corresponding $RLC$-circuit after stopping temporal modulation. (b)--The instantaneous voltage and the power consumed by the resistance $R=1$~$\Omega$. Notice that here, $t_0=\pi/(4\omega)$, $\omega=1$~rad/sec, $R_0=1$~$\Omega$, $A=1$~V and $I_0=1$~A/sec.} 
\label{fig_RLC_circuit_power}
\end{figure}

Modulating the elements in time as expressed in Eq.~(\ref{eq:LtCt}), no energy is reflected back to the source and all the input energy is continuously accumulated in the $LC$ circuit. However, the reactances exchange energy also with the device which modulates their values in time. Thus, we need to consider the power balance and find how much energy is accumulated in the reactive circuit taking into account also the power exchange with the modulating system.   To do this, we assume that the time modulation of the circuit elements stops at a certain time moment ($t_0$) and the circuit inductance and capacitance do not depend on time at later times $t>t_0$. This means that at $t>t_0$ there is no power exchange with the system which modulates the reactances. At $t=t_0$ we  connect a parallel resistance to the $LC$ circuit as shown in Fig.~\ref{fig_RLC_circuit_power}(a), to form a usual $RLC$ circuit with time-invariant elements. We choose the moment $t_0$ at which the inductance and capacitance are both positive and calculate the energy delivered to the resistor during the relaxation time. This energy is equal to the energy which has been accumulated in the time-modulated circuit during the time $0<t<t_0$. The rate of releasing the stored energy depends on the value of the resistance. If it is a small resistance, the accumulated energy is consumed in a short period of time. Let us choose $t_0=\pi/(4\omega)$ as the moment when we stop modulation and energy accumulation. For the $RLC$ circuit in Fig.~\ref{fig_RLC_circuit_power}(a), we can write the second-order differential equation in form
\begin{equation}
LC\frac{d^2i_{\rm L}(t)}{dt^2}+\frac{L}{R}\frac{di_{\rm L}(t)}{dt}+i_{\rm L}(t)=0.
\label{eq:current_diff_eq}
\end{equation}  
Regarding the voltage over the elements, we know that $v(t)=L{di_{\rm L}(t)}/{dt}$. Solving the characteristic equation of the $RLC$ circuit, we obtain the electric current $i_{\rm L}(t)$ as
\begin{equation}
i_{\rm L}(t)=A_1e^{S_1t}+A_2e^{S_2t},
\label{eq_inductance_current}
\end{equation}
where  
\begin{equation}
S_{1,2}=\frac{-L\pm\sqrt{L^2-4R^2LC}}{2RLC}.
\label{eq:eigenvalues}
\end{equation}
In Eq.~(\ref{eq_inductance_current}), $A_1$ and $A_2$ are unknown coefficients which can be found  by imposing the initial conditions, i.e.~the current flowing through the inductance and the voltage over the capacitance should be continuous. In other words,
\begin{equation}
\begin{split}
&i_{\rm L}(t)\bigg|_{t=t_0}=\bigg[\frac{A\cos(\omega t_0)}{2R_0}+I_0t_0\bigg]=\alpha,\cr  
&\frac{di_{\rm L}(t)}{dt}\bigg|_{t=t_0}=\frac{A\cos(\omega t_0)}{L}=\beta.
\end{split}
\label{eq:conditionsi}
\end{equation}
According to Eqs.~(\ref{eq_inductance_current}) and~(\ref{eq:conditionsi}), the coefficients $A_1$ and $A_2$ can be written as
\begin{equation}
\begin{split}
&A_1=\frac{\beta-\alpha S_2}{S_1e^{t_0S_1}-S_2e^{t_0S_1}},\cr 
&A_2=\frac{\beta-\alpha S_1}{S_2e^{t_0S_2}-S_1e^{t_0S_2}}.
\end{split}
\end{equation}
Knowing the electric current $i_{\rm L}(t)$, we can calculate the resistance voltage and finally the instantaneous power as $p(t)=v(t)^2/R$ and the total released energy by integrating the instantaneous power from $t_0$ to infinity. This energy is the one that we can extract and use after stopping the modulation. Note that since at $t_0=\pi/(4\omega)$ the values of $L$ and $C$ are such that $S_1$ and $S_2$ are real and not equal, the $RLC$ circuit is \emph{over-damped}. The time dependence of the instantaneous power is shown in Fig.~\ref{fig_RLC_circuit_power}(b). We find that the energy consumed by the resistance is about $W_{\rm{ext}}\approx0.42$~J. Let us compare this value  with the energy delivered to the matched $LC$ circuit from the power source during the accumulation time from $t=0$ till $t=t_0$, which equals $W_{\rm{del}}\approx 0.64$~J. Hence, the time-modulated load not only accumulated all the incident power but also accepted some power from the system which modulated the two reactances. 
\begin{figure}[t!]
	\centering
	\subfigure[]{\includegraphics[width=0.4\textwidth]{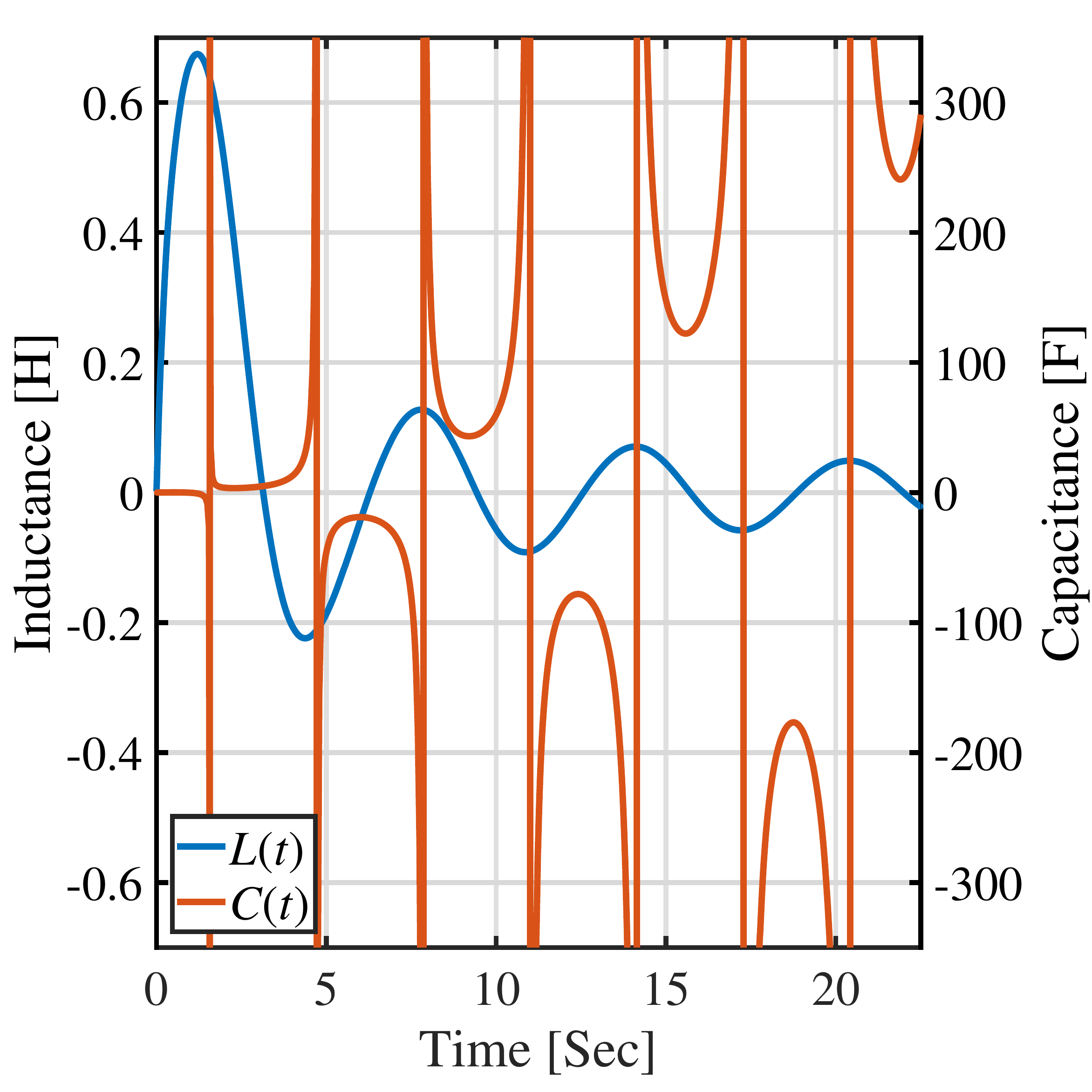}}
	\subfigure[]{\includegraphics[width=0.4\textwidth]{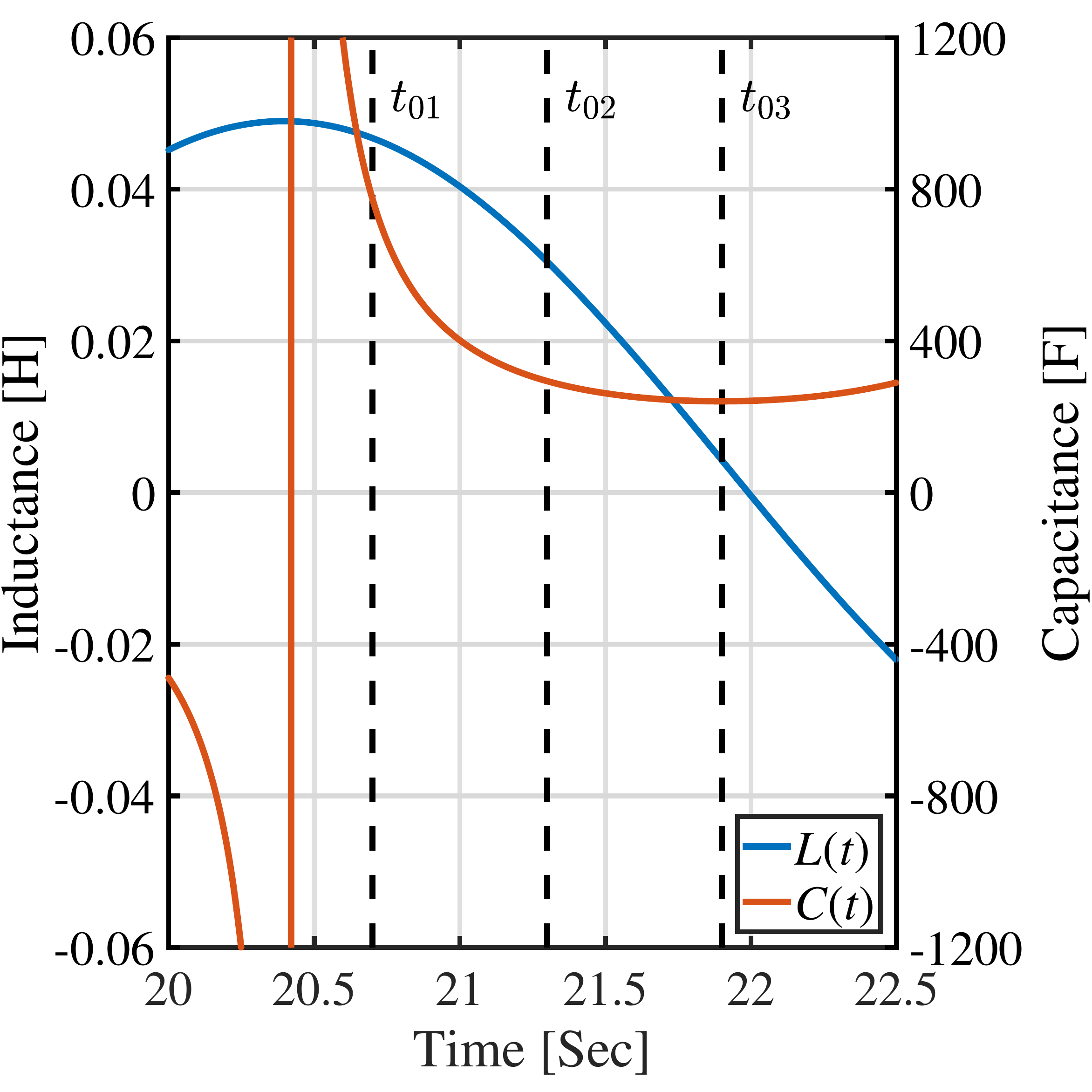}}
	\caption{Time-dependent inductance and capacitance for $R_0=1$~$\Omega$, $\omega=1$~rad/sec, $A=1$~V and $I_0=1$~A/sec.}
	\label{fig:impedance_circuit0}
\end{figure}  

\begin{figure*}[t!]
	\centering
	\subfigure[]{\includegraphics[width=0.4\textwidth]{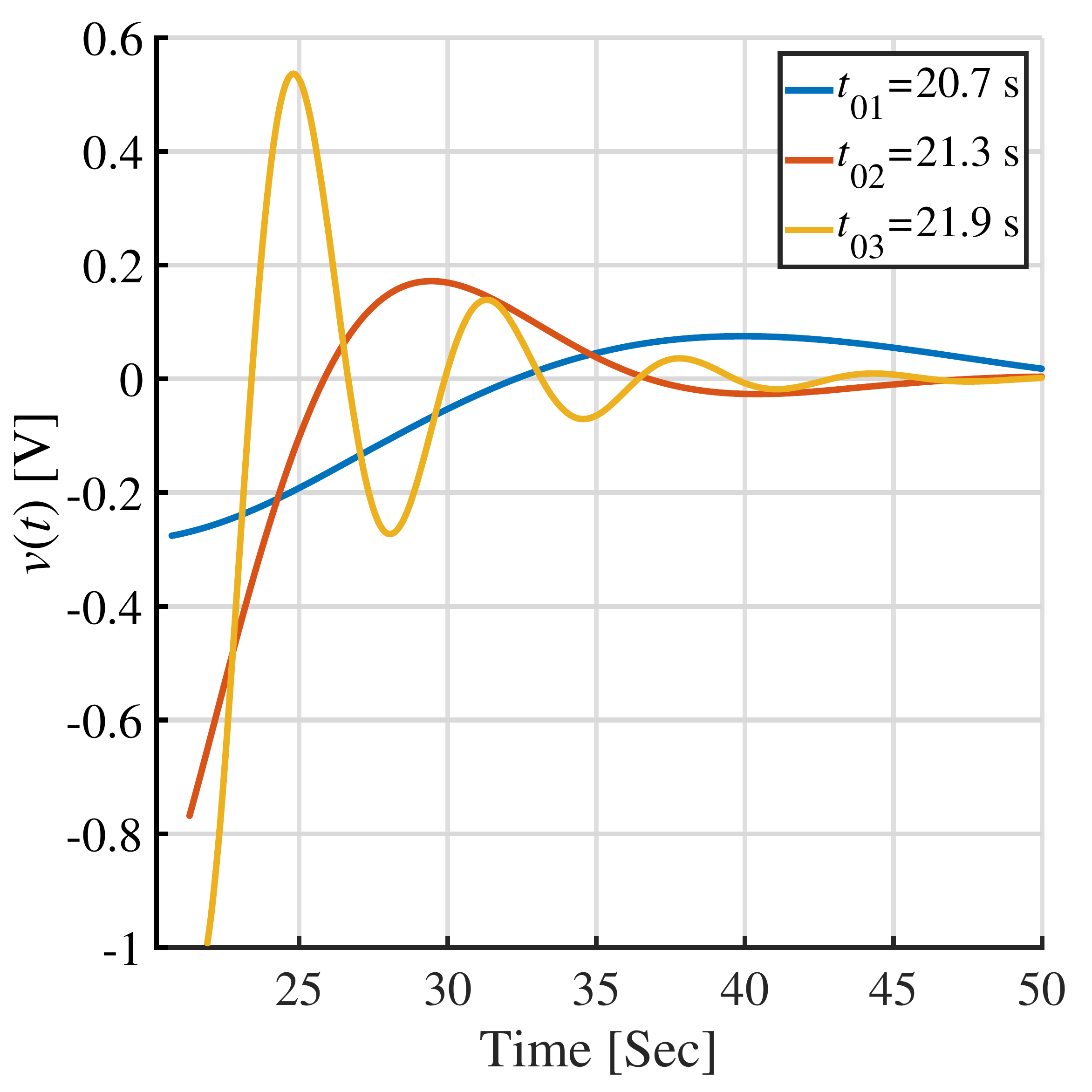}}
	\subfigure[]{\includegraphics[width=0.4\textwidth]{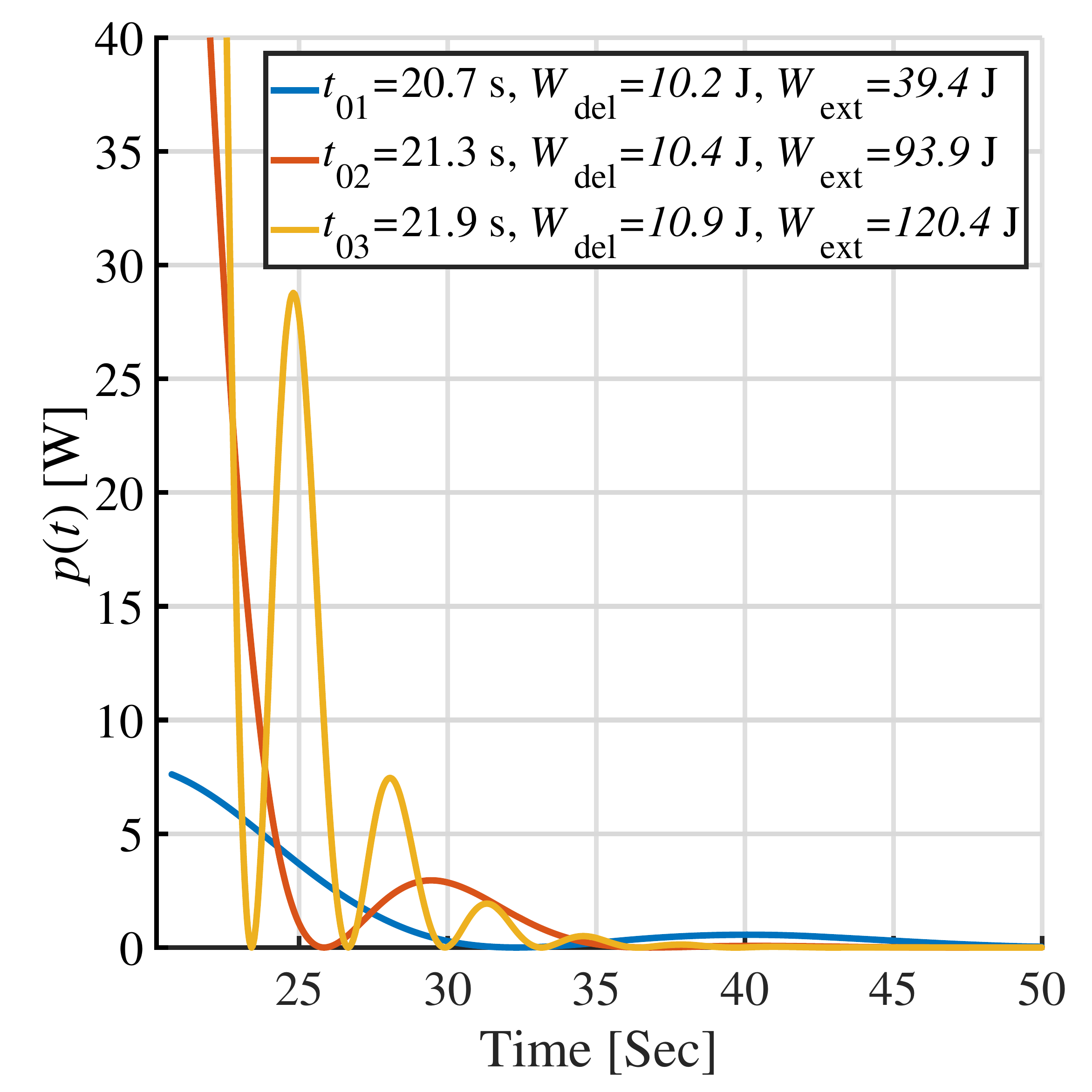}}
	\caption{(a)--The resistance voltage and (b)--the corresponding instantaneous power consumed by the resistance ($R=0.01$~$\Omega$) after the time modulation is stopped at different time moments. Notice that $R_0=1$~$\Omega$, $\omega=1$~rad/sec, $A=1$~V and $I_0=1$~A/sec.}
	\label{fig:impedance_circuit111}
\end{figure*}  

The above example with $t_0=\pi/(4\omega)$ corresponds to a short energy-accumulation time. It is interesting to investigate energy accumulations for longer times. Figure~\ref{fig:impedance_circuit0} shows that near $t=20.5$~s, there is an asymptote for the capacitance function and the inductance is positive. Let us stop the reactance modulation at the following moments: $t_{01}=20.7$~s, $t_{02}=21.3$~s, and $t_{03}=21.9$~s, and connect a $0.01$~$\Omega$ resistance to the time-invariant $LC$ circuit at these moments. We choose such a small resistance value to release the accumulated energy quickly.  Based on  Eq.~(\ref{eq:eigenvalues}), because the value of the inductance is very small and the capacitance is very large,  we expect that $S_{1,2}$ are complex and conjugate of one another: $S_1=S_2^\ast$. In other words, the circuit is \emph{under-damped}. This feature can be seen in Fig.~\ref{fig:impedance_circuit111} where we show the voltage over the resistance and the instantaneous power for the three different scenarios described above. Calculating the released energy, we find that while the delivered energy does not change much in these three cases ($W_{\rm{del}}\approx 10.2,\,10.4$ and $10.9$~J), the energy which is accumulated and then extracted changes dramatically. It is worth noting that the extracted energy can be much larger than the delivered energy, showing that the modulated  $LC$ circuit accepts and accumulates energy also from the modulation  source, in addition to the energy delivered by the source feeding the transmission line. However, stopping modulation and keeping $L$ and $C$ constant in time is only one way of extracting the energy. It is also possible to release the energy without stopping the modulation. We must only change the modulation function. Hence, we can choose a proper modulation function such that all the energy accepted by the $LC$ circuit from the modulation source will return to the modulation source. In other words, an equal exchange of energy happens between the $LC$ circuit  and the modulation source in the accumulating and releasing regimes. In this scenario, the delivered energy becomes equal to the released (extracted) energy.

\section{Conclusions}
\label{sectionconclulast}
In this paper, we have shown that properly time-variant reactive elements can continuously accumulate energy from conventional external time-harmonic sources, without any reflections of the incident power. We have found  the required time dependences of reactive elements and discussed possible  realizations as time-space modulated transmission lines and mixer circuits. Interestingly, there is a conceptual analogy of energy-accumulating reactances and short-circuited transmission lines where the short position is moving, which helps to understand the physical mechanism of energy accumulation and release. We have shown that properly modulating reactances of two connected elements it is in principle possible to engineer any arbitrary time variations of currents induced by time-harmonic sources. The study of the energy balance has revealed that such parametric circuits accept and accumulate power not only from the main power source but also from the pump which modulates the reactances. This is seen from the fact that if we stop energy accumulation at some moment of time and release all the accumulated energy into a resistor, the released energy can be much larger than the energy delivered to the circuit from the primary source.  These energy-accumulation properties become possible if the time variations of the reactive elements are not limited to periodical (usually time-harmonic) functions, but other, appropriate for desired performance, time variations are allowed.

\section*{Acknowledgement} 
The authors would like to thank Dr.~Anu Lehtovuori for useful discussions on circuits with varying parameters.

\end{document}